\documentclass[runningheads]{llncs}
\usepackage[T1]{fontenc}
\usepackage{graphicx}
\usepackage[hyphens]{url}
\usepackage[hidelinks]{hyperref}
\hypersetup{breaklinks=true}
\usepackage{listings}
\usepackage{multirow}
\usepackage{csquotes}
 
\begin{document}
\title{Improving compiler support for SIMD offload using Arm Streaming SVE}

%
\titlerunning{Compiler support for Arm SSVE}
%

\author{\small Mohamed Husain Noor Mohamed, Adarsh Patil\textsuperscript{\textdagger}, \\ Latchesar Ionkov, Eric Van Hensbergen}
\institute{Arm (Austin, USA), \textsuperscript{\textdagger}Arm (Cambridge, UK)}
\authorrunning{MHN. Mohamed, A. Patil et al.}

\maketitle              
\begin{abstract}
\vspace{-2em}
The wider adoption 
of tightly coupled core-adjacent accelerators, such as Arm Scalable Matrix Extension (SME), hinges on lowering software programming complexity.
In this paper, we focus on enabling the use of SME architecture in Streaming Scalable Vector Extension (SSVE) mode for workloads written in C/C++. 
While current compilers optimize loops for all types of SIMD instructions, these techniques primarily target vector units within the core and falter when applied to disaggregated, core-adjacent SIMD accelerators. Our goal is to enable the compiler to automatically generate code for such accelerators only \textit{when profitable}.
 
To this end, we investigate a path towards performant, precise, and repeatable computation offloading through two compiler ecosystems.
We revisit LLVM compiler passes, MLIR transforms and their associated cost models, and heuristics. 
We hope that these insights can provide directions for evolving compiler capabilities towards automatic code generation for this next-generation vector processing paradigm.

\keywords{Auto-vectorization  \and Arm SME \and Arm Streaming SVE}
\end{abstract}

\section{Introduction}

Modern CPUs feature specialized accelerators adjacent to the main core to offload fine-grained computational tasks.
The SME architecture in Armv9 exemplifies this architecture, accelerating matmul and wide vector width computations, providing dual benefits of energy-efficient execution and reduced design complexity for specialization. 
The most versatile way of lowering programming complexity for such architectures is to use compiler ecosystems to automatically offload computations to these accelerators when profitable. 

\begin{figure}[tbh]
	\centering
	\includegraphics[width=\columnwidth]{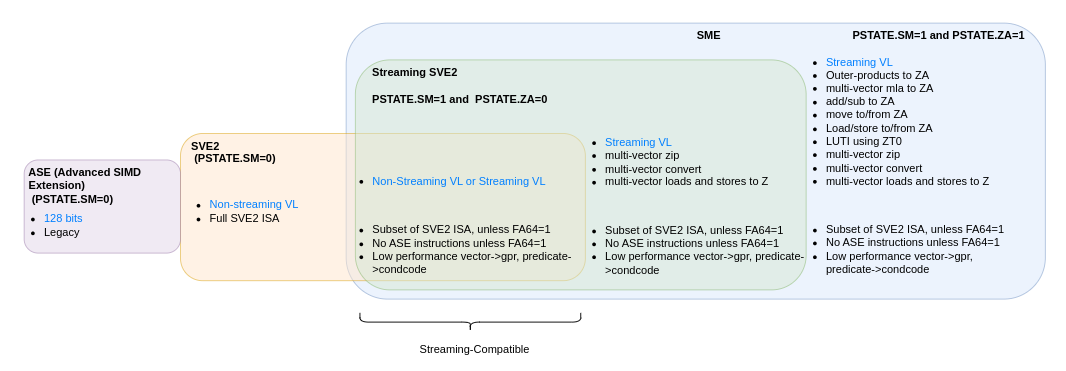}
	\caption{A venn diagram of Arm vector ISA extensions. This study focuses on SVE2 streaming-compatible instructions to utilize the wider vectors of the SME hardware.}
	\label{fig:instructions-venn-diagram}
\end{figure}

\vspace{0.05in}
\noindent \textbf{Streaming SVE (SSVE) mode:}
Arm SME defines SSVE mode \cite{ssve-arm} 
which allows execution of a subset of SVE2 instructions with a flexible, implementation-dependent vector width. This Streaming SVE Vector Length (SVL) can be any power of two in the range of 128-2048 bits. SSVE mode can be enabled by programming the \texttt{PSTATE.SM} field.
The focus of this paper is to enable the use of SSVE mode to perform vector operations, for workloads written in C/C++.
Therefore, we generate SSVE vector pipes, i.e., SVE code running in streaming mode with wider vector width. We do not target Matmul with outer-product and accumulate (MOPA) or Matmul with multi-vector instructions or instructions that use the ZA register storage. Specifically, the target is instructions that are valid in \verb*|PSTATE.SM=1| and \verb*|PSTATE.ZA=0|, commonly referred to as \enquote{streaming compatible instructions} (Fig. \ref{fig:instructions-venn-diagram}.) \cite{sme-arm-arm}

\vspace{0.05in}
\noindent \textbf{SSVE architecture:} 
A potential implementation of SME would support Streaming SVE mode with an SVL as a power of two in the range 128-2048 bits inclusive \cite{briston-sim-eng}.
Each core supports fixed 128-bit vector length NEON SIMD and does not support SVE natively.
Architecturally, these SME units act as part of each core i.e., instructions are sent to the SME unit when streaming mode is enabled on a core. This control data like addresses and offsets are sent over a dedicated control bus. 
Other data are accessed by the SME unit through loads and stores. The SME units have a private L1 cache and a high-bandwidth connection to a shared cache. 
Note that the closest common cache for both the core and the SME unit is the shared cache, i.e. whatever data is written to the memory by the core will be read by the SME unit from the cache and vice versa.

\vspace{0.05in}
\noindent \textbf{Compiler auto-vec limitations:}
Contemporary compilers possess the capability to identify and optimize loops to generate SIMD instructions that execute on vector units located within the core.
In contrast, SSVE architectures have disaggregated  SIMD units outside of the core.
Consequently, the entire infrastructure for estimating profitability of vectorization is relatively simple and ill-suited for architectures implementing streaming SVE.
We revisit the applicability of existing techniques for generating vectorization candidates, transforms, cost models, and heuristics to achieve automatic offload capability in a variety of compilers.

The heuristic of whether to generate SVE instructions vs fixed-width NEON instructions should include the microarchitectural features, e.g., cost of switching into/out of streaming mode, the associated costs of running in streaming mode, performance of fall back path. An incorrect decision can potentially make the performance worse than falling back to fixed-width vector/scalar mode. 

\vspace{0.05in}
\noindent \textbf{Contributions:} \\
(i) We demonstrate compiler limitations in auto-vectorizing for SSVE  offload, using a variety of benchmarks on a SME supported processor (Sec. \ref{sec:motivation}).  \\
(ii) We propose the enhancements necessary for LLVM and MLIR-based compilers to facilitate the generation of SSVE code (Sec. \ref{sec:approach}).






\section{Motivation} \label{sec:motivation}



\textbf{Compiler limitations:} Auto-vectorization is disabled in LLVM for streaming-enabled functions. Using the following experimental flag, 
\begin{lstlisting}[breaklines]
-mllvm -enable-scalable-autovec-in-streaming-mode
\end{lstlisting} 
forces auto-vec which generates SVE code without a streaming-mode switch. Developers must then annotate the function with \verb*|__arm_streaming| to enable streaming mode.
In addition, careful analysis and profiling are needed to ensure: \\ 
(a) \textit{correctness} - the auto-vectorizer's \textit{code-generator} may vectorize using SVE instructions that are not streaming compatible, e.g. scatter/gather, leading to \verb|EXC_BAD_INSTR| \\
(b) \textit{performance} - the auto-vectorizer does not have a \textit{cost-model} to decide profitability of vectorization using SSVE leading to severely degraded performance, potentially worse than scalar execution.

\noindent \textbf{Code-gen performance analysis:} 
We benchmark to assess compiler auto-vectorization capability and architecture performance (config in Table \ref{table:compiler-flags}). We use the \verb|-mcpu| flag to tune the cost model for the vector lengths of our processor architecture.
Recall that the SME unit in the processor offers SIMD vector widths much wider than that of the NEON unit on the core.
Consequently, we expect to see some improvements in performance for certain workloads even with such a rudimentary compilation \cite{llvm-20}.

\begin{table}[h]
\caption{Experimental setup for SSVE benchmarking}
\label{table:compiler-flags}
\begin{tabular}{|c|p{0.9\linewidth}|}
\hline
Clang & \begin{tabular}[c]{@{}l@{}}clang version 19.1.4; Target: arm64\end{tabular} \\ \hline
Platform Arch & Processor with support for 128-bit NEON and 512-bit SVL \\ \hline
SSVE & -O3  -fstrict-aliasing   -mllvm -enable-scalable-autovec-in-streaming-mode \\ \hline
NEON & -O3 -fstrict-aliasing  -fvectorize \\ \hline
Scalar& -O3 -fstrict-aliasing  -fno-vectorize \\ \hline
\end{tabular}
\end{table}


\noindent \textbf{$\succ$ TSCV\_2 suite \cite{tsvc-modified}:} 
(Iterations 100000, Sizes: LEN\_1D 32000, LEN\_2D 256).
Among the 146 loops in the TSVC suite, we observe 

(i) \textit{Performance:} \textit{Baselining to scalar execution,} compiling to SSVE sees a geomean of 0.52x (upto 153x slowdown). Only 53\% of the loops show improved performance with SSVE. Note that compiling to NEON shM  ows a geomean speedup of 1.62x over scalar.
\textit{Baselining to NEON,} SSVE shows a geomean of 0.32x (upto 268x slowdown) with only 22 loops showing better performance with SSVE.

(ii) \textit{SSVE auto-vectorization:} 
Almost 90\% of the loops were auto-vectorized to NEON but only 35\% were auto-vectorized to SSVE by the compiler.
Specifically, we highlight five loops that were autovectorized to SSVE with significant slowdowns compared to NEON and Scalar (s115, s132, s2233, s1281, s443). Across this subset of loops, SSVE shows geomean of 0.14x over NEON and 0.37 over scalar. This clearly highlights the compiler's inadequate cost-model, given that it produced SSVE code even though it led to performance reductions compared to NEON and scalar.


\noindent \textbf{$\succ$ Mandelbrot set \cite{mandel-simd}} 
(Input sizes: 180x135, 360x270, 720x540, 1440x1080) \\
For the C++ source, the compiler SSVE auto-vec failed due to an early exit loop condition. Instead, we used an implementation of the kernel with intrinsics. Across these input sizes, SSVE sees a geomean of 0.35x over NEON. Specifically, we observe that the slowdown decreases linearly with increasing input sizes, since the cost of offload is amortized relative to the execution latency.

\noindent \textbf{$\succ$ SPEC 2017 \cite{spec2017}} We observe a 1.9x slowdown over NEON for the notoriously difficulticult to optimize mcf benchmark (rate mode)  in the SPEC 2017 suite.

\vspace{0.05in}
\noindent \textbf{Motivation summary:}
Although 4x wider SSVE SIMD units offer potential performance gains, most benchmarks show significant slowdowns
The issue largely stems from compilers' inability to generate optimal code. Relying on programmer expertise to diagnose issues or simply isolating acclerator interaction through expertly hand-written libraries limits the use of SME architecture.

\section{Extensions for auto-vectorization to SSVE} \label{sec:approach}
We offer a brief overview of the two approaches (LLVM and MLIR-based) to generate SSVE instructions, highlighting their constraints and limitations.

\subsection{LLVM Approach} \label{sec:llvm}

\textbf{Enhancing the Loop Vectorizer (LV) pass:}
LV widens instructions in loops to operate on multiple consecutive iterations. LV uses a Vectorization Factor (VF) to model the profitability of vectorization over scalar execution, representing the number of vector lanes in the architecture.
For SVE auto-vec the cost model was enhanced to support unknown number of lanes.
Since the vector length is unknown, the compiler needs to ensure that vectorization is beneficial for all vector sizes. To avoid having to check all potential vector lengths for profitability, LV assumes that speedup is constant or strictly increasing with vector size, i.e., code beneficial for vectorizing at VF=4 is also advantageous at VF=16. It therefore performs a single cost analysis for the smallest possible vector size.

\textbf{$\succ$ Implementation details:} To support SVE in LV, \textit{VectorType} was split into \textit{FixedVectorType} and \textit{ScalableVectorType}. 
The parameter \textit{vscale} was introduced to support scalable vector types in LV \cite{vscale}.
A scalable vector type is specified of the form $<vscale \times N \times Ty>$ where `\textit{N}' is the minimum number of elements of type `\textit{Ty}' contained within the vector, and `\textit{vscale}' indicates that the total element count is an integer multiple of `\textit{N}'.
\textit{vscale} is an unknown at compile time and takes a value greater than 0 and in the range specified by \textit{vscale\_range(min[,max])}.
For example, a vector containing an unknown integer multiple of four i32s is represented as $<vscale \times 4 \times i32>$.

VF for profitability analysis is expressed as $<vscale \times N>$. 
Flag \verb*|-mtune=<cpu>| is used to tune for specific vscale based on the microarchitecture. For example \verb*|-mtune=neoverse-v1| uses vscale=2.
If the parameter is not specified, the code generated remains compatible with any \textit{vscale} in the specified \textit{vscale\_range}.

\textbf{$\succ$ Extending the cost-model:} 
The cost model analysis uses the codegen to approximate the cost of any IR instruction when lowered to machine instructions. The cost results are unit-less and the cost number represents the throughput of the machine assuming that all loads hit the cache, all branches are predicted, etc. Cost numbers can be added to compare two or more transformation alternatives.

The current cost model does not take into account that speedups might not scale perfectly linearly and increasing vector lengths using SSVE can have added performance impacts on speedups. The cost-model must account for the performance overheads of running in SSVE mode and the associated non-linear performance scaling for various vector lengths in \textit{vscale\_range(min[,max])} .
In addition, since SSVE only supports a subset of SVE2 instructions, invalid SVE operations, such as scatter-gather, now need to be excluded from the cost-model calculations.

Note that simply changing the instruction costs in the compiler table will not work. The cost of an SVE2 instruction can be different in streaming vs non-streaming mode. Additionally, the cost overheads of SSVE instructions are not associated with individual instructions but are amortized over a basic block. 

\vspace{0.05in}
\noindent \textbf{Enhancing Vectorization Plan (VPlan):} 
A VPlan is a representation of a distinct way to vectorize a loop nest, called a vectorized candidate. This candidate is represented using a hierarchical CFG. VPlan supports estimating the cost and driving the generation of the output IR code. The Hierarchical CFG models the planned control-flow, whose nodes are basic blocks (VPBasicBlock).

The LoopVectorizationPlanner handles the vectorization of a loop or a loop nest. It constructs, optimizes, and discards one or more VPlans, each VPlan modeling a distinct way to vectorize the loop or the loop nest. Once the best VPlan is determined, including the best VF and UF, this VPlan drives the generation of the output IR.
A VPBasicBlock holds a sequence of zero or more VPRecipes which model the cost and generation of the output IR instructions. 

\textbf{$\succ$ Extending the cost model:} VPlan cost modeling totals each instruction's cost in the loop body to estimate a plan's cost. It calls cost() on VPlans, which recurses into VPBasicBlocks and into VPRecipes. Most cost() methods for individual recipes currently call CostModel.getInstructionCost which uses the look-up of the tables for instruction costs.
We propose adding costs to VPlans that employ streaming SSVE. The cost() method for an SSVE VPlan's cost() method should be instantiated with a static cost to account for streaming mode overheads before calculating instruction costs. 

\subsection{MLIR Approach} \label{sec:mlir}
\textbf{Enhancing Polygeist \cite{polygeist-pact21}:} We describe extensions to Polygeist - a C/C++ compilation flow using MLIR \cite{mlir}. 
Polygeist currently auto-vectorizes loops for fixed width vector NEON instructions only and cannot generate SVE.

\begin{figure}
	\centering
	\includegraphics[width=0.9\columnwidth]{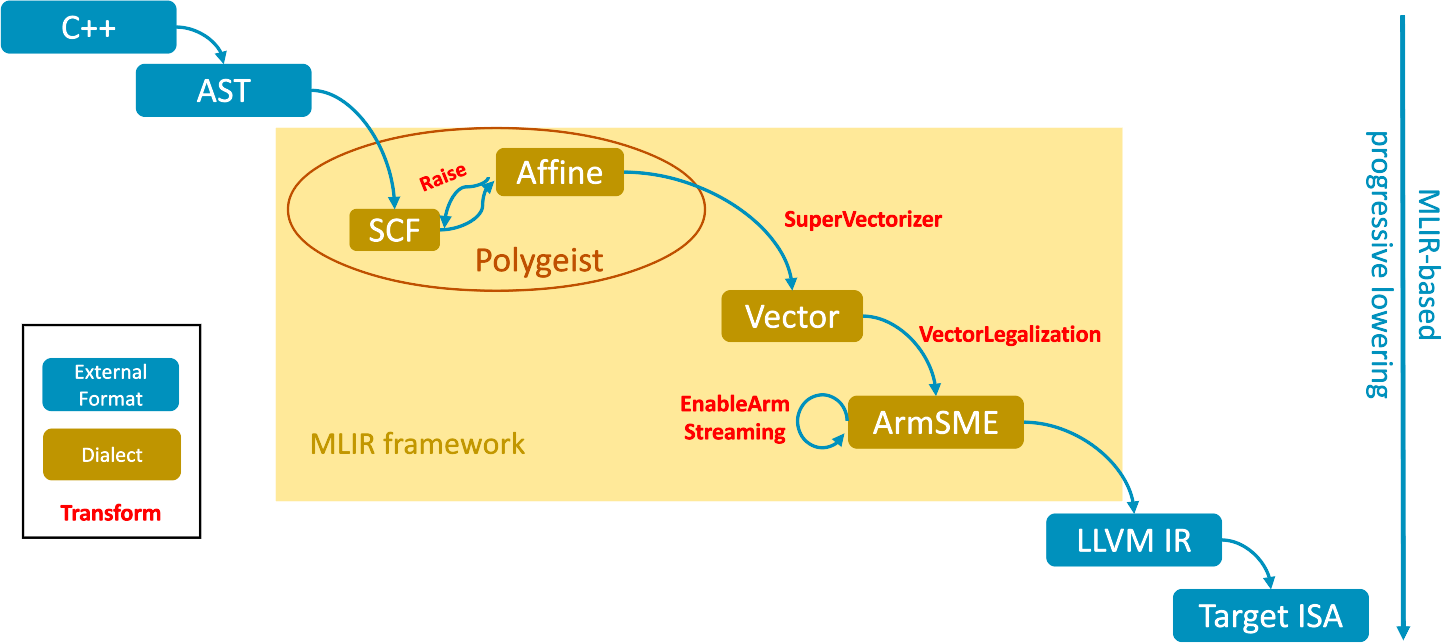}
	\caption{The proposed extensions for lowering for generating SSVE. Within the MLIR framework (yellow box), only the red circle is currently present in Polygeist.}
	\label{fig:proposed-mlir-lowering}
\end{figure}

\textbf{$\succ$ Implementation details:} Fig \ref{fig:proposed-mlir-lowering} illustrates the proposed compilation pipeline, which extends Polygeist, to generate SSVE code. Polygeist currently generates and optimizes the code only at the Affine level. The proposed pipeline enhances existing transforms and dialects to generate SSVE as detailed below.

\textit{SuperVectorizer transform:} 
We propose to apply the SuperVectorizer transform to lower the optimized Affine dialect to the Vector dialect. 
A \enquote{super-vector} is loosely defined as a vector type that is a multiple of a \enquote{good} vector size so the hardware can efficiently implement a set of high-level primitives. Loops and operations are emitted that operate on those super-vector shapes, which are later lowered to hardware-specific vector sizes. Currently, the SuperVectorization transform currently does not generate scalable vectors, and doing so requires also adding a cost-model heuristic in the transformation.

\textit{Vector dialect:} Vector dialect currently supports representing scalable vectors. As an example, 
a fixed-width vector represented as $vector<4 \times 8 \times 128 \times f32>$ lowers to  $!llvm<[4 \times [ 8 \times < 128 \times float >]]>$ in LLVM IR. \\
a scalable vector representation represented as $vector<4 \times 8 \times [128] \times f32>$ lowers to  $!llvm<[4 \times [ 8 \times < vscale \times 128 \times float >]]>$ in LLVM IR.

\textit{ArmSME and ArmSVE dialect:}
The ArmSVE dialect currently does not flag streaming compatible SVE instructions. As SSVE is an ArmSME feature, we suggest extending the ArmSME dialect with SSVE operations, using the same sematics as in ArmSVE. ArmSME  does provide an operation to query the streaming length \verb*|arm_sme.streaming_vl|. Extending the ArmSME dialect allows us to use its transforms to generate streaming SVE (see below).

\textit{VectorLegalization and EnableArmStreaming transforms:}
These transforms are part of the ArmSME dialect. They currently have very limited support for the SSVE operations that are in scope for this study.
The \textit{VectorLegalization} transform legalizes vector operations for ArmSME, focusing on SME tiling. This transform needs extensions for SSVE operations. A similar legalization transform is performed when lowering vector operations in ArmSVE via LegalizeVectorStorage.
\textit{EnableArmStreaming} transform enables streaming mode by adding arm\_streaming or arm\_locally\_streaming functions. The LLVM backend will emit `smstart sm' / `smstop sm' [4] around calls to streaming functions.

\subsection{Architectural considerations for the SSVE cost-model} \label{sec:cost-model}
To calibrate the auto-vectorizers cost model for streaming-SVE, we outline the architectural consideration of the SME unit that affect the performance of the generated code.

\vspace{0.5em}
\noindent \textbf{Synchronizations:} Synchronizations between core and the SME unit stall execution. Minimizing these will improve performance of the generated code.

\begin{itemize}
    \item \textit{GPR and FPR synchronization:} In streaming mode, the floating-point operations are dispatched to the SME unit which has a private L1 cache. When both that SME unit and the scalar core access the same cache lines on the stack, the memory-ordering hardware must stall on one side until the other retires its access.
To reduce the probability of such events during streaming execution, the AArch64 back-end in LLVM can insert a tunable pad (specified by \verb|-aarch64-stack-hazard-size| flag) between stack objects predominantly accessed through GPRs and those accessed through FPRs. 
The padding pushes FPR-only data to the middle of the frame and GPR-only data at the edges.

This heuristic cannot isolate variable-length arrays, argument spill slots, or objects referenced by both register classes \cite{stack-hazard-size}. Micro-benchmarking on the SME platform shows that the residual stall penalty is 17\% for a single streaming loop, grows roughly linearly to 61\% for 100 concatenated loops, and then mostly saturates.

\item \textit{Predicates synchronization:}
The predicate registers are used to select the active lanes in the SME unit. Both the core and the SME unit can produce predicates. If the core uses the produced predicate from the SME, e.g. the load-store instructions in streaming mode, synchronization of predicates produced on the core adds tens of cycles of latency. In code generation, to avoid this synchronization replacing the predicates with the vectors registers and placing the predicates usage as far as possible will help improve the performance.
\end{itemize}

\noindent \textbf{Memory interactions:} Memory interactions between the SME unit occur through the shared cache. Minimizing such interactions in the generated code will improve performance.

\begin{itemize}
\item \textit{Load-Store Region Table (LSRT) Hazards:} The SME unit relies on the core's MMU for address translation for each load/store instruction.  The core maintains the LSRT for the physical address and is synchronized to the SME unit using LSRT sync-up packets. 
Each time the SME unit requests a load or store, if the region is unallocated, a new LSRT region gets created and synchronized with the SME unit.
The core is also responsible for verifying that the transfer is successful by checking translation faults, alignment faults, and watchpoints.  The address translation in the LSRT is instantiated at 4KB granularity.

A core access is stalled until the completion of all the SME unit accesses to the same region are completed and visible by the coherency. Similarly, the access cannot be sent to the SME unit if it conflicts with a core access in the same region. The core uses the LSRT to detect these hazards between the core requests and the SME unit. Accesses within the same region are monitored by the LSRT table. Each LSRT entry implements counters to track these hazards at an aligned 1KB granularity regions. LSRT hazards lead to performance slowdowns.

To avoid these hazards, the compiler can generate spills and fills of both scalar/vector data accessing the same region. Secondly, the compiler can keep 1KB aligned regions in the stack if it is accessed both by the core and the SME unit, and at least one executes write accesses.

We observe that the QPSK modulation kernel in Arm RAL \cite{arm-ral} compiled for SSVE causes shows 6$\times$ slowdown over NEON. Using the above technique to avoid the LSRT hazard improves the performance of the kernel.


\item \textit{Scaling behavior:} The performance gains of streaming SVE over regular NEON or scalar is dependent on the input length. For smaller input lengths, offloading to the
SME unit incurs higher costs (warming up SME caches, memory access characteristics of the loop, chances of LSRT stalls due to the source and destination memory mapping to the same 1KB LSRT region, etc.). We note that most loops tend to show a performance knee, a point after which SSVE is more performant than NEON, amortizing the costs of offload to the SME unit
Compiler cost-model decision-making heuristics can be conservative, defaulting to NEON, as the consequences of an inefficient offload can degrade performance.

\item \textit{Prefetcher:} The SME unit prefetcher is based on strided and nested access patterns and is less capable of tracking arbitrary access efficiently. Arbitrary access can pollute training and prevent useful prefetching.The compiler can revert to NEON if pointers or data-dependent addresses are used in the loop.

\item \textit{Gather loads and scatter stores:} These instructions are unavailable in streaming SVE. Regular strided accesses can be converted to regular load/store SSVE
instructions using ZIP/UZIP instructions. However, random stride cause performance slowdowns, Therefore, the general guidance for compiler code generation is to execute loops with gather/scatter operations using NEON instructions.

\end{itemize}

\section{Conclusions and call to action}
Compiler auto-vectorizers need updating to support core-adjacent offload vector computation. It is crucial to factor in the architectural characteristics such as mode switch/synchronization costs and data access via accelerator caches to assess offload profitability.
More generally, this paper motivates compiler researchers to enable a transition to this more efficient paradigm of compute offloading, especially as application stacks evolve to be architecture-agnostic. 

\bibliography{refs}
\end{document}